\documentclass[reprint,superscriptaddress,amsmath,amssymb,aip,jap]{revtex4-1}

\usepackage{lipsum}
\usepackage{graphicx}
\usepackage{dcolumn}
\usepackage{bm}
\usepackage{hyperref}
\usepackage{tabularx}
\usepackage[version=4]{mhchem}
\usepackage{ulem}

\usepackage{xcolor}
\usepackage{soul}

\sethlcolor{white}

\newcommand{\XYq}{\mathrm{X}^q_\mathrm{Y}}

\begin{document}


\title{Quick-start guide for first-principles modelling of point defects in crystalline materials}

\author{Sunghyun Kim}
\affiliation{Department of Materials, Imperial College London, Exhibition Road, London SW7 2AZ, UK}
\affiliation{Department of Materials Science and Engineering, Yonsei University, Seoul 03722, Korea}

\author{Samantha N. Hood}
\affiliation{Department of Materials, Imperial College London, Exhibition Road, London SW7 2AZ, UK}

\author{Ji-Sang Park}
\affiliation{Department of Physics, Kyungpook National University, Daegu 41566, Korea}

\author{Lucy D. Whalley}
\affiliation{Department of Materials, Imperial College London, Exhibition Road, London SW7 2AZ, UK}


\author{Aron Walsh}
\email{a.walsh@imperial.ac.uk}
\affiliation{Department of Materials, Imperial College London, Exhibition Road, London SW7 2AZ, UK}
\affiliation{Department of Materials Science and Engineering, Yonsei University, Seoul 03722, Korea}

\date{\today}

\begin{abstract}
\hl{Defects influence the properties and functionality of all crystalline materials. For instance, point defects participate in electronic (e.g. carrier generation and recombination) and optical (e.g. absorption and emission) processes critical to solar energy conversion. Solid-state diffusion, mediated by the transport of charged defects, is used for electrochemical energy storage.} First-principles calculations of defects based on density functional theory have been widely used to complement, and even validate, experimental observations. In this `quick-start guide', we discuss the best practice in how to calculate the formation energy of point defects in crystalline materials and analysis techniques appropriate to probe changes in structure and properties \hl{relevant across energy technologies}. 
\end{abstract}

\maketitle



\textit{`The perfect crystal is one of the idealisations commonly found in theoretical physics and science fiction'.}\cite{Stoneham1983}
Defects are present in all crystals -- they can be detrimental to device performance or can be beneficial for use in a wide variety of processes, including:
\begin{itemize}
\item \textit{optical}, e.g. the colour centre \ce{V_{Br}} in CsBr;
\item \textit{chemical}, e.g. the catalytic centre \ce{Li_{Mg}} in MgO;
\item \textit{mechanical}, e.g. hardening of Fe using C;
\item \textit{electrical}, e.g doping by \ce{V_O} in ZnO.
\end{itemize}
Computational techniques allow us to investigate properties of point defects at a level of detail that is often difficult to access via experiments.
It is possible to isolate the behaviour of particular defects and predict their spectral signatures and physical effects.
One challenge is that much of the infrastructure for materials modelling is built on translational symmetry (e.g. Bloch wave functions). 
Defects break the periodicity of a crystal,
and their accurate description is a continuing endeavour for materials modelling.

Following our quick-start guide on interfaces,\cite{park2018quick} this primer is for researchers starting to work on first-principles simulations of imperfect crystals.
\hl{ 
Point defects play an important role across emerging technologies including thermoelectrics (doping density), batteries (ion diffusion rates), electrocatalysis (active sites), and solar cells (radiative efficiency). 
Accurate predictions of defect processes are critical to our understanding of current materials and the exploration of new compounds with enhanced performance. 
}

\begin{figure}
\includegraphics[width=7.0cm]{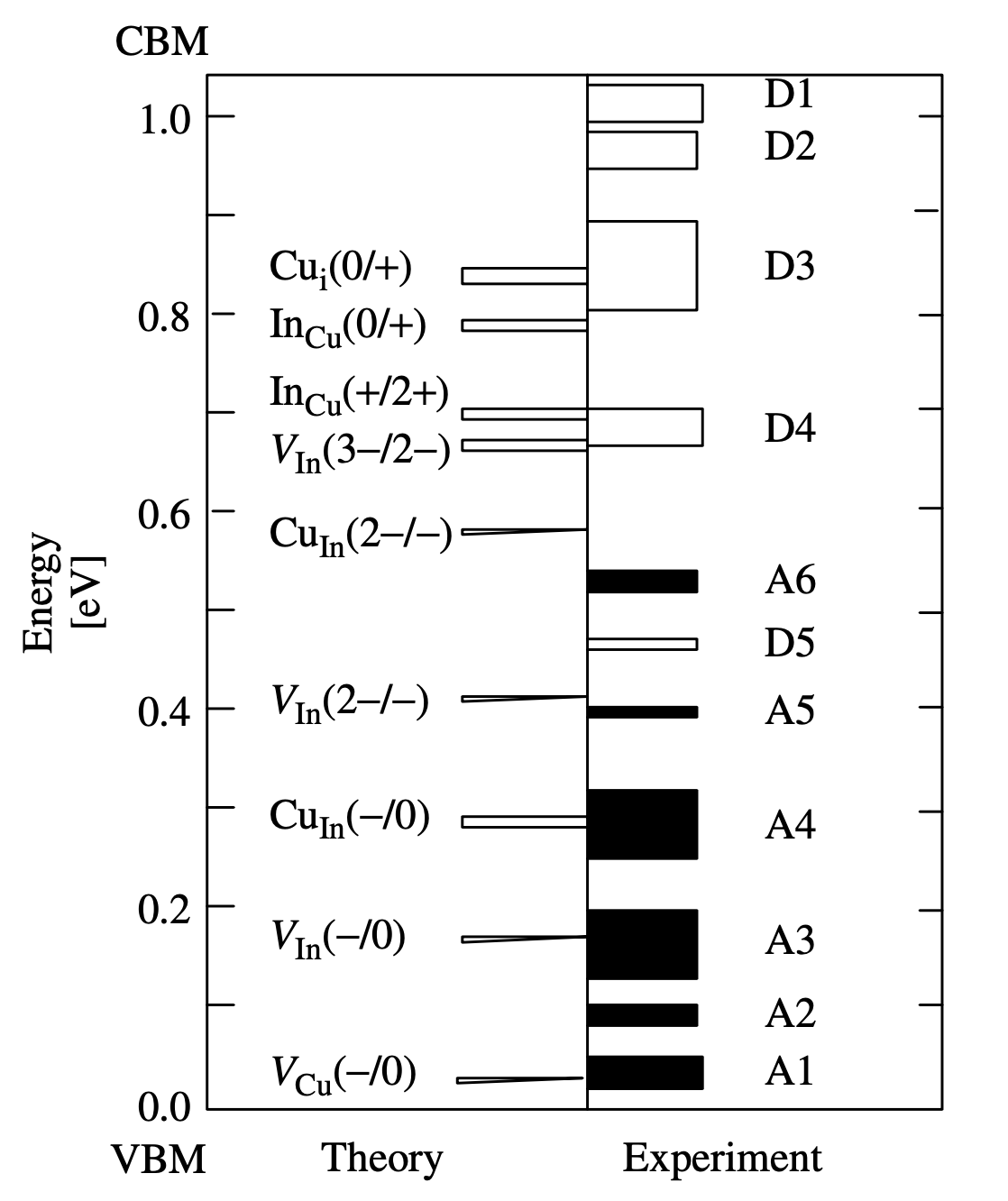}
\caption{\label{f-cigs} 
Intrinsic defect levels in \ce{CuInSe2} with respect to the valence (VBM) and conduction (CBM) bands that have been measured and calculated from first-principles. 
The height of the histogram columns on the right side represents the spread in experimental data. 
Reproduced with permission from Ref. \onlinecite{zhang1998defect}.
} 
\end{figure}

\begin{figure*}
\includegraphics[width=13.0cm]{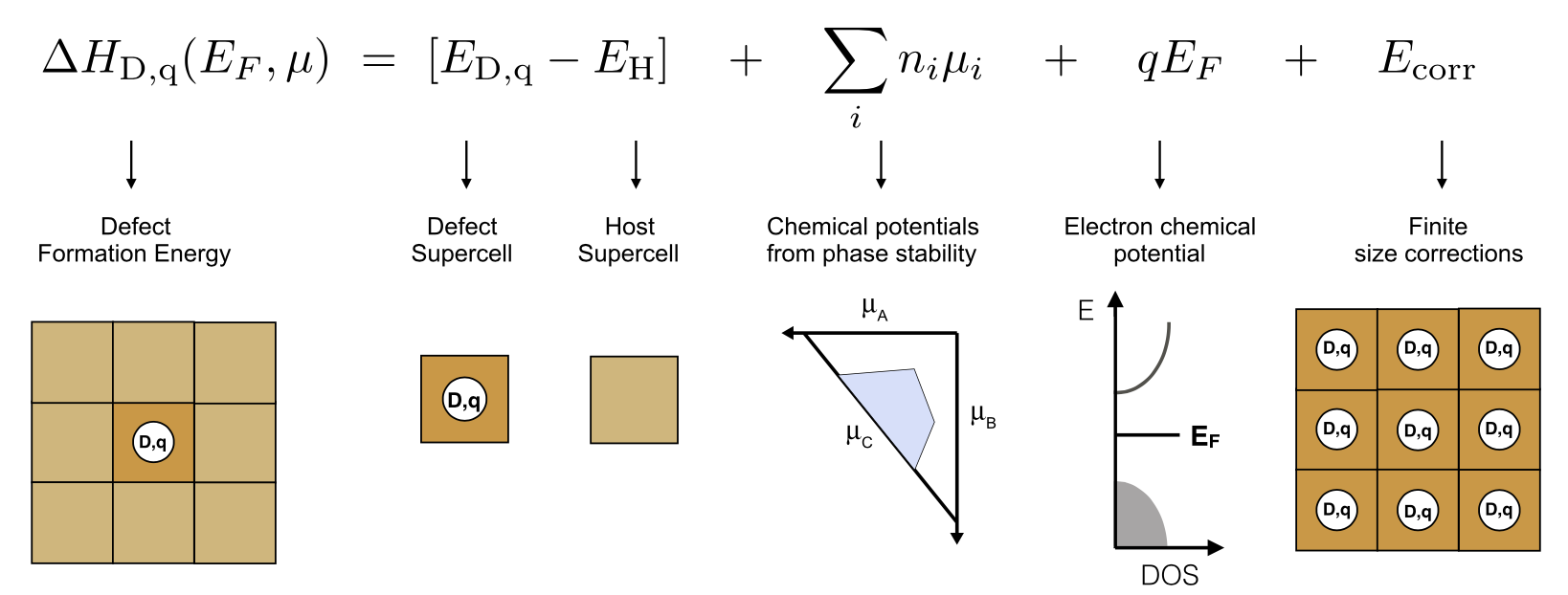}
\caption{\label{f-def} 
Illustration of the terms required to compute the charged defect formation energy as function of the atomic and electronic chemical potentials. Reproduced with permission from Ref. \onlinecite{Goyal2017}.
} 
\end{figure*}

\section*{1. Defect Notation}

A defect is often referred to according to its spectroscopic signature.
An anion vacancy with a trapped electron in an ionic crystal may absorb light in the visible range, making the transparent host material colourful; an $F$ center (\textit{Farbe} means colour in German).
In electrical measurements, such as deep level transient spectroscopy (DLTS), defects are labeled in order of their energy levels.
For example, $D1$ and $D2$ for electron donors and $A1$ and $A2$ for electron acceptors, as shown for the case of the chalcopyrite semiconductor \ce{CuInSe2} in Fig. \ref{f-cigs}.
Matching atomic models with spectroscopic signals is of considerable importance in physics and chemistry of materials.

A widely used notation for the atomic defect models is Kr\"oger-Vink. \cite{Kroger:1974gb}
A defect is represented by \ce{X_Y},
where X is the species occupying the atomic site Y.
A vacancy and an interstitial site are denoted by $V$ and $i$, respectively.
The relative charge of a defect can be represented by a superscript of $ ^{\times}$, $ ^{\bullet}$ and $'$ for a neutral, positive and negative charges, respectively.
A numerical notation\cite{IUPAC_notation} of the charge state ($q$) is popular in the recent literature. 
For example, the negatively-charged B-on-Si is represented by \ce{B^{-1}_{Si}}, 
and the neutral Si self-interstitial is represented by \ce{Si^{0}_{i}}.
When more than one symmetrically inequivalent site exists, an additional subscript of the Wyckoff position or the point group symmetry of the site can be added to distinguish distinct species.

\section*{2. Equilibrium Concentrations}

The equilibrium concentration of defects ($n_d$) at a fixed temperature and pressure is given by the density that minimises the free energy:
\begin{equation}
    n_d = N_\mathrm{site} g \exp \left(-\frac{\Delta G_f}{k_\mathrm{B} T} \right),
\end{equation}
where $N_\mathrm{site}$ and $g$ denote the number of available sites of the defect in the unit volume and the degeneracy of the defect, respectively.
$\Delta G_f$ is the Gibbs free energy of formation of the defect, $k_\mathrm{B}$ is the Boltzmann constant, and $T$ is temperature.
This can be further decomposed into contributions from enthalpy ($H$) and vibrational entropy ($S$) as follows:
\begin{equation}
    n_d = N_\mathrm{site} g \exp \left( - \frac{\Delta H_f}{k_\mathrm{B} T} \right) 
        \exp \left( \frac{\Delta S}{k_\mathrm{B}}\right),
\end{equation}
The enthalpy change dominates under standard conditions and is easier to compute, therefore vibrational terms are often neglected.
\hl{The inclusion of vibrational entropy terms does become important for the description of gaseous phases (e.g. oxygen gas) or for defect formation at elevated temperatures.}\cite{gillan1983entropy}

The formation energy of a defect is given as:
\begin{equation}\label{eq:E_form}
    \Delta H_{f}=\Delta E + \sum n_{i} \mu_{i} + q E_{F} + E_\mathrm{corr},
\end{equation}
where $\Delta E$ is a change in the total energy due to the formation of the defect. 
The second term takes into account the energy cost to exchange $n_i$ atoms of kind $i$ with their reservoir chemical potential $\mu_i$
(e.g. this term for $ \XYq $ is $ \mu_{\mathrm{X}} - \mu_{\mathrm{Y}} $).
$q$ and $E_F$ denote the charge state of the defect and the Fermi level, respectively.
The correction term ($E_\mathrm{corr}$) will be discussed in the next section.
These terms are illustrated in Fig. \ref{f-def}.
The typical magnitudes of first, second, and third terms in Equation \ref{eq:E_form} are a few eV,
while the value of the forth term is usually $<$ 1 eV.

\begin{figure}
\includegraphics[width=8.3cm]{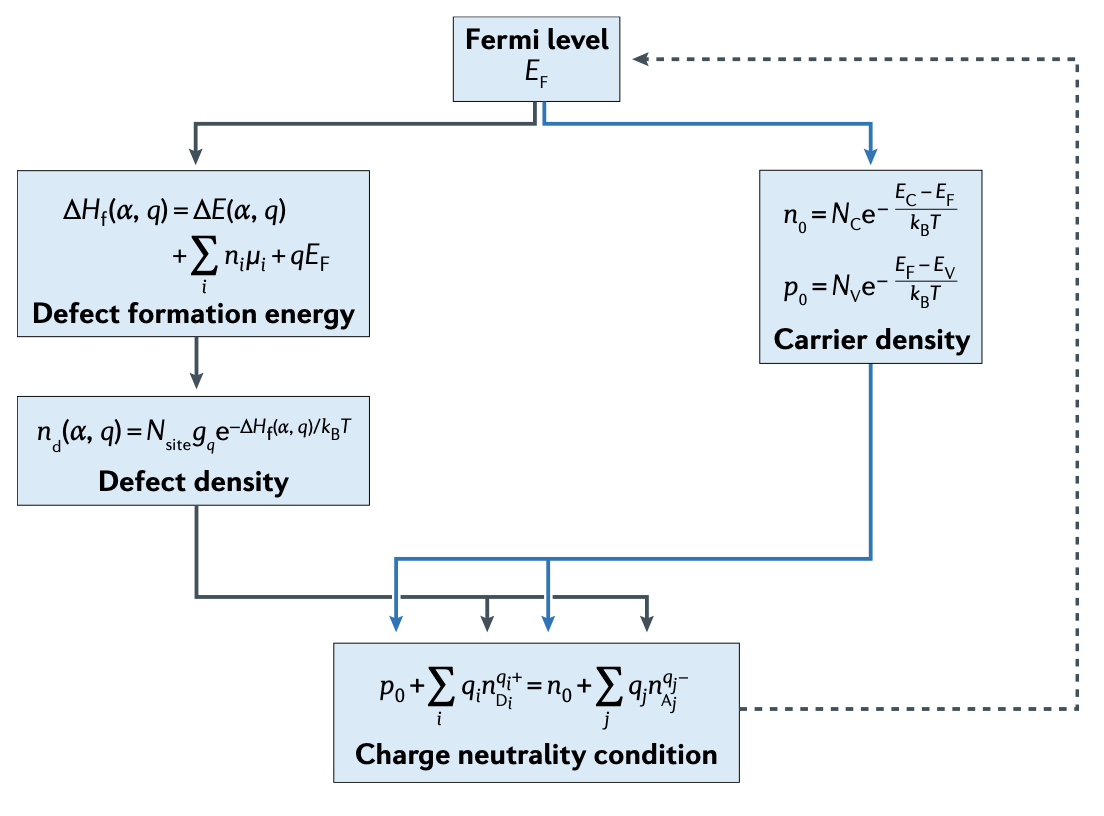}
\caption{\label{fig-sc} 
Illustration of a self-consistent treatment of charged defect populations in a crystalline host material. Reproduced with permission from Ref. \onlinecite{park2018point}.
} 
\end{figure}

Since a macroscopic crystal should be charge neutral overall, the concentrations of electrons $n_0$, holes  $p_0$, positively-charged donors $n_{D_i}^{q_i}$, and negatively-charged acceptors $n_{A_j}^{q_j}$ must satisfy electroneutrality:\cite{park2018point}
\begin{equation}
    p_{0}+\sum_{i}q_{i}n^{q_{i}}_{D_i}=n_{0}+\sum_{j}q_{j}n^{q_{j}}_{A_j}.
\end{equation}
The equilibrium population of charge carriers are given by:
\begin{eqnarray}
n_{0} = N_{C} e^{-\frac{E_{C}-E_{F}}{k_\mathrm{B} T}}, \\
p_{0} = N_{V} e^{-\frac{E_{F}-E_{V}}{k_\mathrm{B} T}}, 
\end{eqnarray}
where $N_{C}$ and $N_{V}$ are the effective density of states of the conduction ($E_{C}$) and valence ($E_{V}$) bands, respectively.
Since the formation energy of charged defects depends on the Fermi level (Equation 3), which in turn depends on the population of charged defects, the concentrations must be solved self-consistently as illustrated in Fig. \ref{fig-sc}. 

\section*{3. Practical calculations}

Several approaches exist for calculating defect formation energies. 
Those based on embedding potentials (a dilute defect in a host matrix) offer several advantages but remain technically challenging to setup and analyse.\cite{xie2017demonstration,buckeridge2015determination} 
For first-principles approaches, including density functional theory (DFT), 
the supercell method is by far the most widely employed.\cite{Freysoldt:2014ej}

With the supercell method, periodic boundary conditions remain in place, but an expanded repeat unit is employed. 
This repeat unit should be large enough so that the host material is well described and the periodic interactions between repeating defects can be corrected.
For example, if a cubic unit cell with lattice spacing ($a$) is expanded using a matrix of:
\begin{equation}
 \begin{bmatrix}
   2  &  0 & 0       \\
   0  & 2 & 0        \\
   0  & 0 & 2        \\
\end{bmatrix} 
\end{equation}
there will be a spacing of 2$a$ between repeating defects along each axis.

Vacancies and substitutions can be introduced by simply removing atoms and replacing atoms, respectively.  
Interstitial sites require further care because of the larger configurational space. 
Candidate sites in the crystal to form interstitials can be assigned in a way that resembles well-known structure motifs such as tetrahedra or octahedra.\cite{Broberg2018}
\texttt{PyLada} adapts a scheme based on Voronoi tessellation.\cite{Goyal2017}
The range of possible charge states is inferred from the oxidation states of atoms.
For example, the charge state of \ce{Sn_{Zn}} can be 0 or +2 which correspond to Sn(II) and Sn(IV), respectively.
\hl{However, the chemistry occurring at defect sites can be unexpected, e.g. cation-cation bonding, so a search over a wide range of charge states is often necessary to identify the accessible configurations.}

\hl{
Within DFT, the choice of exchange-correlation functional is an important factor to consider. 
Defects often result in localised wavefunctions, where electron correlation effects are intensified.
If a given functional performs well (e.g. crystal structure and band gap) for the perfect crystal, it does not guarantee that it will provide an accurate description of the associated defects. 
The most notable failures have been for local or semi-local functionals (e.g. the local density approximation, LDA).
For example, an oxygen vacancy in TiO$_2$ results in delocalised donor states in LDA, while a more sophisticated description using DFT$+U$ or a hybrid non-local functional correctly describes the localised reduction of two Ti(IV) to Ti(III)}.\cite{morgan2009polaronic}

We will now breakdown Equation 3 into its components and discuss how to compute the four terms in turn.

\subsection{Term 1: Raw defect formation energy}

To obtain $\Delta E$, the standard operating procedure is as follows:
\begin{enumerate}
\item Fully optimise the crystal structure of the host material in its \textit{primitive unit cell}.
\item Create a \textit{supercell} expansion that is as close to cubic as possible, which minimises anisotropy in defect-defect interactions.
\item The total energy of the pristine \textit{supercell} becomes the $E_H$ reference for defect formation energies.
\item Introduce a defect of your choosing and optimise the structure under \textit{constant volume} conditions keeping the lattice vectors fixed.  
\item The total energy of the optimised defective \textit{supercell} is  $E_{D,q}$.
\item The raw defect formation energy in Equation 3 is the difference of two total energy calculations, $\Delta E = E_{D,q} - E_H$.
\end{enumerate}

One common mistake is to assume an incorrect spin state. 
Even if a host material is non-magnetic, the ground state of a defect may require spin-polarisation, e.g. in an open-shell singlet or triplet configuration.\cite{sokol2010oxygen} 

\subsection{Term 2: Atomic chemical potentials}

The raw defect formation energies are not meaningful as they do not represent balanced reactions (atoms may have been created or annihilated).  
The chemical potentials $\mu_i$ account for the exchange of species with their environment.
In practice, a range of environments are possible.
In some cases we may try to mirror an experiment, e.g. an environment of oxygen gas at a particular temperature or pressure. More generally, we treat $\mu_i$ as a parameter than can vary over the accessible phase space of the material. 
The standard operating procedure, \hl{following Zhang and Northrup,}\cite{zhang1991chemical} is as follows:
\begin{enumerate}
\item Calculate the total energy of the standard states of each element found in your host material (e.g. Zn metal and oxygen gas for ZnO).
\item Calculate the total energy of all possible secondary phases that could form (e.g. ZnO and \ce{Al2O3} for a \ce{ZnAl2O4} host).
\item Solve the accessible chemical potential region considering these boundaries. Any point in this stability field can be used safely in Equation 3.
\end{enumerate}
For multi-component materials, the associated simultaneous equations become cumbersome to solve. 
One freely available package developed for this purpose is \texttt{CPLAP}.\cite{buckeridge2014automated}

\subsection{Term 3: Fermi level}

In semiconductors and dielectrics, the electronic chemical potential (Fermi level) can vary between the valence and conduction bands, depending on the doping regime and history of a sample. 
Here, $qE_F$ represents the cost of exchanging electrons or holes with the host and is therefore proportional to the defect charge. 
It has become standard to present defect formation energies as a function of a parametric $E_F$ in the range [0,$E_g$], where $E_g$ is the bandgap of the host compound.
In reality, the full range will not always be accessible.
The equilibrium Fermi level can be solved,\hl{ as shown by Aradi \textit{et a}l}\cite{aradi2001ab}, conveniently using a software package such as \texttt{SC-FERMI}.\cite{buckeridge2019equilibrium}

\begin{table*}[ht]
\begin{tabular*}{\textwidth}{c @{\extracolsep{\fill}} ll}
    Code & Purpose & Correction Scheme \\ \hline \hline
    PyLada~\cite{Goyal2017} & Automate point defect calculations & Lany-Zunger \\ 
    CoFFEE~\cite{Naik2018} & Electrostatic corrections for charged defect calculations & FNV \\
    PyDEF~\cite{Pean2017} & Defect formation energies using VASP & Lany-Zunger \\
    PyCDT~\cite{Broberg2018} & Facilitate high-throughput  DFT defect calculations & FNV and KO \\
    sxdefectalign~\cite{FNV2009} &  Point defects in bulk within SPHinX~\cite{Marquardt2012} & FNV \\
    sxdefectalign2d~\cite{Freysoldt2018interface} & Point defects at surfaces and interfaces within SPHinX & FNV \\ 
    CPLAP\cite{buckeridge2014automated} & Calculation of stable chemical potential ranges  & .. \\ 
    SC-Fermi\cite{buckeridge2019equilibrium} &  Determine the Fermi level based on defect formation energies & .. \\
    CarrierCapture.jl~\cite{Kim:2020bx} & Calculate non-radiative carrier capture by anharmonic defects & ..
\end{tabular*}
\caption{A selection of packages available to assist in processing first-principles defect calculations. }\label{table:code}
\end{table*}

\subsection{Term 4: Charged defect corrections}

The treatment of charged defects is a longstanding issue for periodic boundary conditions due to the long-range nature of the Coulomb interaction.
There are two issues to resolve. 
Firstly, charged defects are able to interact with their periodic images. 
Secondly, a homogeneous ``jellium" background charge is introduced to enforce charge neutrality and a convergent Coulomb energy.
These issues result in a shift in the average electrostatic potential of the supercell and in $E_{D,q}$.
Large supercells are optimal to reduce these errors; however, we are often limited by computational cost and available resources. 
A number of correction schemes have been developed that result in the term $E_{corr}$. Some are discussed below and for a more complete description of these issues, we refer the reader elsewhere.\cite{durrant2018,Vinichenko2017}

\begin{enumerate}
\item The \textit{Leslie-Gillan} correction \cite{Leslie1985} models a point charge $q$ interacting with its periodic images through an isotropic dielectric medium. 
This correction takes a simple analytic form that depends on the charge state $q$, static dielectric constant $\varepsilon$, separation between images $L$ and the Madelung constant $\alpha_m$ characteristic of the lattice.

\item The \textit{Makov-Payne} correction\cite{Makov1995} includes an additional term to account for higher-order multipoles:
\begin{equation}
    E^\mathrm{MP} = \frac{q^2\alpha_{m}}{2\varepsilon L} + qQL^{-3}.
\end{equation}
An issue associated with this approach is in determining the quadrupole moment $Q$. 

\item The \textit{Lany-Zunger} correction\cite{Lany2009} combines the Makov-Payne correction, including a procedure for calculating $Q$, with potential alignment to correct for the shift in electrostatic potential. 

\item The \textit{Freysoldt, Neugebauer and van de Walle} (FNV) method\cite{Freysoldt2009} models the defect charge as a Gaussian distribution. 
The difference between the electrostatic potential of the charged defect and  perfect bulk supercells, calculated far from the defect, is aligned with the defect model potential. 

\item \textit{Kumagai and Oba} (KO) extended the FNV method  using atomic site potentials combined with the Gaussian charge model for an anisotropic dielectric medium.\cite{Kumagai2014} 
\end{enumerate}
Such schemes were initially developed for use with three-dimensional crystal with homogeneous dielectric screening.
Recent work has extended these methods to two-dimensional\cite{Freysoldt2018,Komsa2013} and one-dimensional\cite{Kim2014} materials.

There is currently no standardised approach to defect charge corrections, 
which can lead to a spread in calculated defect formation energies in the literature, and predicted defect densities that differ by orders of magnitude.
FNV and KO have become the most widely used methods.
These are implemented in several computational tools such as \texttt{PyLada} and  \texttt{sxdefectalign}, see Table 1.
Developing an efficient scheme to account for microscopic effects and anisotropy remains an active area of research.\cite{durrant2018,Vinichenko2017}

As a side note, for shallow defects where the valence or conduction bands become occupied, an additional band filling correction can be required to obtain results in the dilute limit.\cite{Lany:2008gk}

\section*{Frontiers of defect modelling}

\subsection*{Carrier capture and recombination}

Our previous discussion was limited to an equilibrium description of defects and charge states (e.g. for a crystal in the dark at a certain temperature).
Either following a pump electromagnetic pulse or under steady-state illumination, the kinetics of carrier (electron and hole) capture by defects becomes important.

\begin{figure}
\includegraphics[width=6cm]{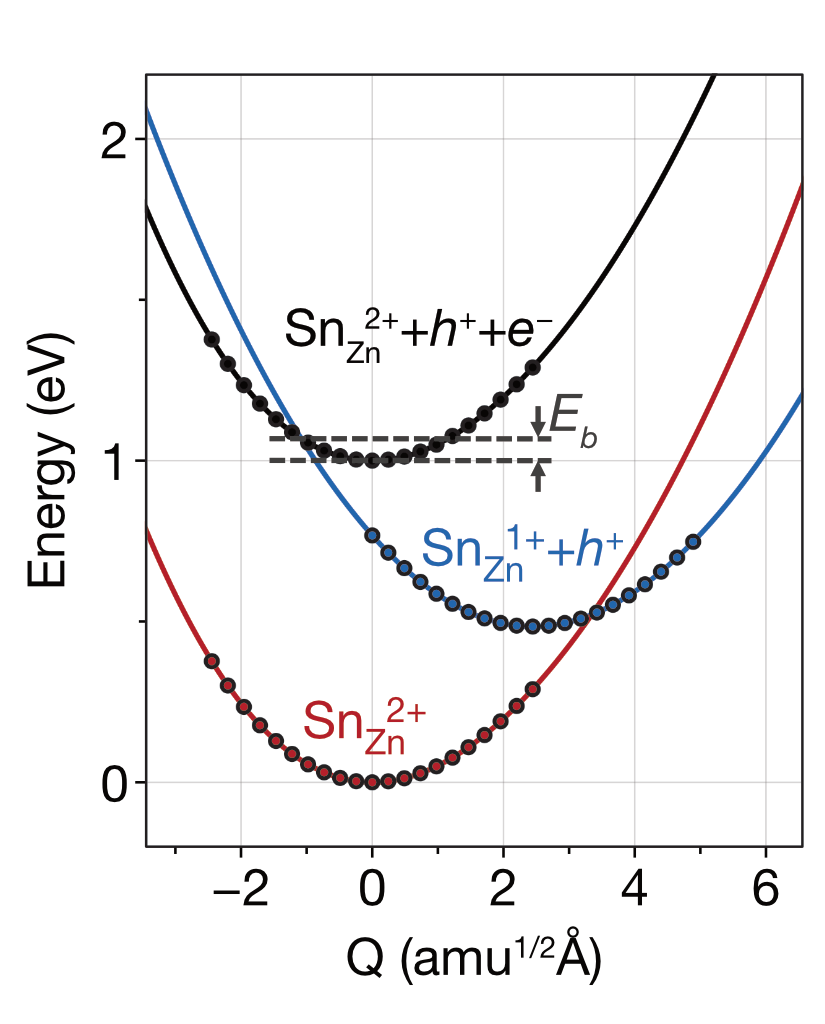}
\caption{\label{f-cc}
Configuration coordinate diagram for \ce{Sn_{Zn}} in \ce{Cu2ZnSnSe4}.
It describes the non-radiative electron-hole recombination process: 
 $\ce{Sn_{Zn}^{2+} <=>[h\nu]
 Sn_{Zn}^{2+} + e^- + h^+ <=>[-\hbar\omega] Sn_{Zn}^{+} + h^+ <=>[-\hbar\omega]  Sn_{Zn}^{2+}}$.
Reproduced with permission from Ref. \onlinecite{Kim:2020sc}.}
\end{figure}

A defect may capture a delocalised free carrier with the aid of electron-phonon coupling.
As the charge is transferred from the delocalised state to the localised state around the defect, the local atomic configuration is rearranged.
Such processes can be described in the framework of a configurational coordinate (CC) diagram. 
The CC is usually in the form of a one-dimensional pathway between two local minimum structures as shown in Fig. \ref{f-cc}.

To describe such non-radiative defect processes, both the nuclear and electron wave functions must be adequately described.
Several schemes have been proposed to calculate the cross-sections and rates of carrier capture based on first-principles simulations.\cite{Shi:2012iv,Alkauskas:2014kk,Kim:2019hp}
The current schemes require a significant amount of researcher expertise and computational resource.
Further developments are required to develop reliable and robust procedures suitable for general applications. 

\subsection*{Dynamics and transport}
Materials engineering often requires a certain spatial distribution of defects. 
To achieve a desirable profile of defect species in a sample, one needs to understand atomic diffusion.  
However, diffusion is a rare event compared to typical vibrational frequencies (1--10 THz).
Standard \textit{ab initio} molecular dynamics simulations can not properly capture diffusion kinetics due to the limited time and length scales that are accessible. 
To overcome such difficulties, statistical methods such transition state theory (TST) \cite{Vineyard:1957iy} can be employed.
The nudged elastic band (NEB) \cite{Henkelman:2000ez} method has been developed to find a minimum energy path and the energy barrier at the saddle point, which can be used to paramaterise kinetic models. 

Real diffusion processes may consist of multiple barriers, even in relatively simple structure types such as zincblende.\cite{yang2015first}
The role of excited states in crystals with defect-mediated mass transport, for example as found in halide perovskite solar cells\cite{kim2018large,walsh2018taking}, is one area that requires further development.

\subsection*{Automation and databases}
Defect studies are demanding on human time due to the large number of individual calculations and corrections involved, as previously outlined. 
There are a growing number of publicly available packages to assist with pre- or post-processing (see Table \ref{table:code}).
Powered by the rapidly expanding computational capacity and the automation frameworks such as \texttt{AiiDA},\cite{pizzi2016aiida}
databases for first-principles calculations including modelling of defects are being developed, which will
offer the opportunity for to gain insights 
when combined with statistical analysis and machine learning models.
Indeed, a proof-of-concept regression model was recently reported to described defects in Cd chalcogenides.\cite{mannodi2020machine} 

\section*{Conclusion}
We have discussed common considerations and procedures for simulating point defects in bulk materials.
First-principles modelling can provide both qualitative and quantitative descriptions of properties of crystalline solids including thermodynamic potentials and spectroscopic signatures.
Emerging renewable energy technologies are already benefiting from first-principles modelling of point defects, complementing experimental characterisation.
The prospect of developing large defect databases, with the aid of automation, offers an opportunity to extract valuable correlations and insight towards \textit{defect-engineered} materials.

\acknowledgments
We thank Se\'{a}n Kavanagh for a careful reading of the manuscript, and Prashun Gorai for useful suggestions.
The research was funded by the Royal Society and the EU Horizon2020 Framework (STARCELL, grant no. 720907).
Additional support was received from the Faraday Institution (grant no. FIRG003) and the EPSRC (EP/L01551X/1).


\bibliography{abbreviated.bib}

\begin{thebibliography}{46}%
\makeatletter
\providecommand \@ifxundefined [1]{%
 \@ifx{#1\undefined}
}%
\providecommand \@ifnum [1]{%
 \ifnum #1\expandafter \@firstoftwo
 \else \expandafter \@secondoftwo
 \fi
}%
\providecommand \@ifx [1]{%
 \ifx #1\expandafter \@firstoftwo
 \else \expandafter \@secondoftwo
 \fi
}%
\providecommand \natexlab [1]{#1}%
\providecommand \enquote  [1]{``#1''}%
\providecommand \bibnamefont  [1]{#1}%
\providecommand \bibfnamefont [1]{#1}%
\providecommand \citenamefont [1]{#1}%
\providecommand \href@noop [0]{\@secondoftwo}%
\providecommand \href [0]{\begingroup \@sanitize@url \@href}%
\providecommand \@href[1]{\@@startlink{#1}\@@href}%
\providecommand \@@href[1]{\endgroup#1\@@endlink}%
\providecommand \@sanitize@url [0]{\catcode `\\12\catcode `\$12\catcode
  `\&12\catcode `\#12\catcode `\^12\catcode `\_12\catcode `\%12\relax}%
\providecommand \@@startlink[1]{}%
\providecommand \@@endlink[0]{}%
\providecommand \url  [0]{\begingroup\@sanitize@url \@url }%
\providecommand \@url [1]{\endgroup\@href {#1}{\urlprefix }}%
\providecommand \urlprefix  [0]{URL }%
\providecommand \Eprint [0]{\href }%
\providecommand \doibase [0]{http://dx.doi.org/}%
\providecommand \selectlanguage [0]{\@gobble}%
\providecommand \bibinfo  [0]{\@secondoftwo}%
\providecommand \bibfield  [0]{\@secondoftwo}%
\providecommand \translation [1]{[#1]}%
\providecommand \BibitemOpen [0]{}%
\providecommand \bibitemStop [0]{}%
\providecommand \bibitemNoStop [0]{.\EOS\space}%
\providecommand \EOS [0]{\spacefactor3000\relax}%
\providecommand \BibitemShut  [1]{\csname bibitem#1\endcsname}%
\let\auto@bib@innerbib\@empty
\bibitem [{\citenamefont {Stoneham}(1983)}]{Stoneham1983}%
  \BibitemOpen
  \bibfield  {author} {\bibinfo {author} {\bibfnamefont {A.~M.}\ \bibnamefont
  {Stoneham}},\ }\href {\doibase 10.1016/0378-5963(83)90040-5} {\bibfield
  {journal} {\bibinfo  {journal} {Appl. Surf. Sci.}\ }\textbf {\bibinfo
  {volume} {14}},\ \bibinfo {pages} {249} (\bibinfo {year} {1983})}\BibitemShut
  {NoStop}%
\bibitem [{\citenamefont {Park}\ \emph
  {et~al.}(2018{\natexlab{a}})\citenamefont {Park}, \citenamefont {Jung},
  \citenamefont {Butler},\ and\ \citenamefont {Walsh}}]{park2018quick}%
  \BibitemOpen
  \bibfield  {author} {\bibinfo {author} {\bibfnamefont {J.-S.}\ \bibnamefont
  {Park}}, \bibinfo {author} {\bibfnamefont {Y.-K.}\ \bibnamefont {Jung}},
  \bibinfo {author} {\bibfnamefont {K.~T.}\ \bibnamefont {Butler}}, \ and\
  \bibinfo {author} {\bibfnamefont {A.}~\bibnamefont {Walsh}},\ }\href@noop {}
  {\bibfield  {journal} {\bibinfo  {journal} {J. Phys. Energy}\ }\textbf
  {\bibinfo {volume} {1}},\ \bibinfo {pages} {016001} (\bibinfo {year}
  {2018}{\natexlab{a}})}\BibitemShut {NoStop}%
\bibitem [{\citenamefont {Zhang}\ \emph {et~al.}(1998)\citenamefont {Zhang},
  \citenamefont {Wei}, \citenamefont {Zunger},\ and\ \citenamefont
  {Katayama-Yoshida}}]{zhang1998defect}%
  \BibitemOpen
  \bibfield  {author} {\bibinfo {author} {\bibfnamefont {S.}~\bibnamefont
  {Zhang}}, \bibinfo {author} {\bibfnamefont {S.-H.}\ \bibnamefont {Wei}},
  \bibinfo {author} {\bibfnamefont {A.}~\bibnamefont {Zunger}}, \ and\ \bibinfo
  {author} {\bibfnamefont {H.}~\bibnamefont {Katayama-Yoshida}},\ }\href@noop
  {} {\bibfield  {journal} {\bibinfo  {journal} {Phys. Rev. B}\ }\textbf
  {\bibinfo {volume} {57}},\ \bibinfo {pages} {9642} (\bibinfo {year}
  {1998})}\BibitemShut {NoStop}%
\bibitem [{\citenamefont {Goyal}\ \emph {et~al.}(2017)\citenamefont {Goyal},
  \citenamefont {Gorai}, \citenamefont {Peng}, \citenamefont {Lany},\ and\
  \citenamefont {Stevanovic}}]{Goyal2017}%
  \BibitemOpen
  \bibfield  {author} {\bibinfo {author} {\bibfnamefont {A.}~\bibnamefont
  {Goyal}}, \bibinfo {author} {\bibfnamefont {P.}~\bibnamefont {Gorai}},
  \bibinfo {author} {\bibfnamefont {H.}~\bibnamefont {Peng}}, \bibinfo {author}
  {\bibfnamefont {S.}~\bibnamefont {Lany}}, \ and\ \bibinfo {author}
  {\bibfnamefont {V.}~\bibnamefont {Stevanovic}},\ }\href {\doibase
  https://doi.org/10.1016/j.commatsci.2016.12.040} {\bibfield  {journal}
  {\bibinfo  {journal} {Comput. Mater. Sci.}\ }\textbf {\bibinfo {volume}
  {130}},\ \bibinfo {pages} {1 } (\bibinfo {year} {2017})}\BibitemShut
  {NoStop}%
\bibitem [{\citenamefont {Kr{\"o}ger}(1974)}]{Kroger:1974gb}%
  \BibitemOpen
  \bibfield  {author} {\bibinfo {author} {\bibfnamefont {F.}~\bibnamefont
  {Kr{\"o}ger}},\ }\href@noop {} {\emph {\bibinfo {title} {{The Chemistry of
  Imperfect Crystals}}}}\ (\bibinfo  {publisher} {North-Holland},\ \bibinfo
  {address} {Amsterdam},\ \bibinfo {year} {1974})\BibitemShut {NoStop}%
\bibitem [{\citenamefont {Connelly}\ \emph {et~al.}(2005)\citenamefont
  {Connelly}, \citenamefont {Damhus}, \citenamefont {Hartshorn},\ and\
  \citenamefont {Hutton}}]{IUPAC_notation}%
  \BibitemOpen
  \bibinfo {editor} {\bibfnamefont {N.~G.}\ \bibnamefont {Connelly}}, \bibinfo
  {editor} {\bibfnamefont {T.}~\bibnamefont {Damhus}}, \bibinfo {editor}
  {\bibfnamefont {R.~M.}\ \bibnamefont {Hartshorn}}, \ and\ \bibinfo {editor}
  {\bibfnamefont {A.~T.}\ \bibnamefont {Hutton}},\ eds.,\ \href@noop {} {\emph
  {\bibinfo {title} {Nomenclature of Inorg. Chem.}}}\ (\bibinfo  {publisher}
  {The Royal Society of Chemistry},\ \bibinfo {year} {2005})\BibitemShut
  {NoStop}%
\bibitem [{\citenamefont {Gillan}\ and\ \citenamefont
  {Jacobs}(1983)}]{gillan1983entropy}%
  \BibitemOpen
  \bibfield  {author} {\bibinfo {author} {\bibfnamefont {M.}~\bibnamefont
  {Gillan}}\ and\ \bibinfo {author} {\bibfnamefont {P.}~\bibnamefont
  {Jacobs}},\ }\href@noop {} {\bibfield  {journal} {\bibinfo  {journal} {Phys.
  Rev. B}\ }\textbf {\bibinfo {volume} {28}},\ \bibinfo {pages} {759} (\bibinfo
  {year} {1983})}\BibitemShut {NoStop}%
\bibitem [{\citenamefont {Park}\ \emph
  {et~al.}(2018{\natexlab{b}})\citenamefont {Park}, \citenamefont {Kim},
  \citenamefont {Xie},\ and\ \citenamefont {Walsh}}]{park2018point}%
  \BibitemOpen
  \bibfield  {author} {\bibinfo {author} {\bibfnamefont {J.~S.}\ \bibnamefont
  {Park}}, \bibinfo {author} {\bibfnamefont {S.}~\bibnamefont {Kim}}, \bibinfo
  {author} {\bibfnamefont {Z.}~\bibnamefont {Xie}}, \ and\ \bibinfo {author}
  {\bibfnamefont {A.}~\bibnamefont {Walsh}},\ }\href@noop {} {\bibfield
  {journal} {\bibinfo  {journal} {Nat. Rev. Mater.}\ ,\ \bibinfo {pages} {1}}
  (\bibinfo {year} {2018}{\natexlab{b}})}\BibitemShut {NoStop}%
\bibitem [{\citenamefont {Xie}\ \emph {et~al.}(2017)\citenamefont {Xie},
  \citenamefont {Sui}, \citenamefont {Buckeridge}, \citenamefont {Catlow},
  \citenamefont {Keal}, \citenamefont {Sherwood}, \citenamefont {Walsh},
  \citenamefont {Scanlon}, \citenamefont {Woodley},\ and\ \citenamefont
  {Sokol}}]{xie2017demonstration}%
  \BibitemOpen
  \bibfield  {author} {\bibinfo {author} {\bibfnamefont {Z.}~\bibnamefont
  {Xie}}, \bibinfo {author} {\bibfnamefont {Y.}~\bibnamefont {Sui}}, \bibinfo
  {author} {\bibfnamefont {J.}~\bibnamefont {Buckeridge}}, \bibinfo {author}
  {\bibfnamefont {C.~R.~A.}\ \bibnamefont {Catlow}}, \bibinfo {author}
  {\bibfnamefont {T.~W.}\ \bibnamefont {Keal}}, \bibinfo {author}
  {\bibfnamefont {P.}~\bibnamefont {Sherwood}}, \bibinfo {author}
  {\bibfnamefont {A.}~\bibnamefont {Walsh}}, \bibinfo {author} {\bibfnamefont
  {D.~O.}\ \bibnamefont {Scanlon}}, \bibinfo {author} {\bibfnamefont {S.~M.}\
  \bibnamefont {Woodley}}, \ and\ \bibinfo {author} {\bibfnamefont {A.~A.}\
  \bibnamefont {Sokol}},\ }\href@noop {} {\bibfield  {journal} {\bibinfo
  {journal} {Phys. Stat. Sol. (a)}\ }\textbf {\bibinfo {volume} {214}},\
  \bibinfo {pages} {1600445} (\bibinfo {year} {2017})}\BibitemShut {NoStop}%
\bibitem [{\citenamefont {Buckeridge}\ \emph {et~al.}(2015)\citenamefont
  {Buckeridge}, \citenamefont {Catlow}, \citenamefont {Scanlon}, \citenamefont
  {Keal}, \citenamefont {Sherwood}, \citenamefont {Miskufova}, \citenamefont
  {Walsh}, \citenamefont {Woodley},\ and\ \citenamefont
  {Sokol}}]{buckeridge2015determination}%
  \BibitemOpen
  \bibfield  {author} {\bibinfo {author} {\bibfnamefont {J.}~\bibnamefont
  {Buckeridge}}, \bibinfo {author} {\bibfnamefont {C.~R.~A.}\ \bibnamefont
  {Catlow}}, \bibinfo {author} {\bibfnamefont {D.}~\bibnamefont {Scanlon}},
  \bibinfo {author} {\bibfnamefont {T.}~\bibnamefont {Keal}}, \bibinfo {author}
  {\bibfnamefont {P.}~\bibnamefont {Sherwood}}, \bibinfo {author}
  {\bibfnamefont {M.}~\bibnamefont {Miskufova}}, \bibinfo {author}
  {\bibfnamefont {A.}~\bibnamefont {Walsh}}, \bibinfo {author} {\bibfnamefont
  {S.}~\bibnamefont {Woodley}}, \ and\ \bibinfo {author} {\bibfnamefont
  {A.}~\bibnamefont {Sokol}},\ }\href@noop {} {\bibfield  {journal} {\bibinfo
  {journal} {Phys. Rev. Lett.}\ }\textbf {\bibinfo {volume} {114}},\ \bibinfo
  {pages} {016405} (\bibinfo {year} {2015})}\BibitemShut {NoStop}%
\bibitem [{\citenamefont {Freysoldt}\ \emph {et~al.}(2014)\citenamefont
  {Freysoldt}, \citenamefont {Grabowski}, \citenamefont {Hickel}, \citenamefont
  {Neugebauer}, \citenamefont {Kresse}, \citenamefont {Janotti},\ and\
  \citenamefont {Van~de Walle}}]{Freysoldt:2014ej}%
  \BibitemOpen
  \bibfield  {author} {\bibinfo {author} {\bibfnamefont {C.}~\bibnamefont
  {Freysoldt}}, \bibinfo {author} {\bibfnamefont {B.}~\bibnamefont
  {Grabowski}}, \bibinfo {author} {\bibfnamefont {T.}~\bibnamefont {Hickel}},
  \bibinfo {author} {\bibfnamefont {J.}~\bibnamefont {Neugebauer}}, \bibinfo
  {author} {\bibfnamefont {G.}~\bibnamefont {Kresse}}, \bibinfo {author}
  {\bibfnamefont {A.}~\bibnamefont {Janotti}}, \ and\ \bibinfo {author}
  {\bibfnamefont {C.~G.}\ \bibnamefont {Van~de Walle}},\ }\href@noop {}
  {\bibfield  {journal} {\bibinfo  {journal} {Rev. Mod. Phys.}\ }\textbf
  {\bibinfo {volume} {86}},\ \bibinfo {pages} {253} (\bibinfo {year}
  {2014})}\BibitemShut {NoStop}%
\bibitem [{\citenamefont {Broberg}\ \emph {et~al.}(2018)\citenamefont
  {Broberg}, \citenamefont {Medasani}, \citenamefont {Zimmermann},
  \citenamefont {Yu}, \citenamefont {Canning}, \citenamefont {Haranczyk},
  \citenamefont {Asta},\ and\ \citenamefont {Hautier}}]{Broberg2018}%
  \BibitemOpen
  \bibfield  {author} {\bibinfo {author} {\bibfnamefont {D.}~\bibnamefont
  {Broberg}}, \bibinfo {author} {\bibfnamefont {B.}~\bibnamefont {Medasani}},
  \bibinfo {author} {\bibfnamefont {N.~E.}\ \bibnamefont {Zimmermann}},
  \bibinfo {author} {\bibfnamefont {G.}~\bibnamefont {Yu}}, \bibinfo {author}
  {\bibfnamefont {A.}~\bibnamefont {Canning}}, \bibinfo {author} {\bibfnamefont
  {M.}~\bibnamefont {Haranczyk}}, \bibinfo {author} {\bibfnamefont
  {M.}~\bibnamefont {Asta}}, \ and\ \bibinfo {author} {\bibfnamefont
  {G.}~\bibnamefont {Hautier}},\ }\href {\doibase
  https://doi.org/10.1016/j.cpc.2018.01.004} {\bibfield  {journal} {\bibinfo
  {journal} {Comput. Phys. Commun.}\ }\textbf {\bibinfo {volume} {226}},\
  \bibinfo {pages} {165 } (\bibinfo {year} {2018})}\BibitemShut {NoStop}%
\bibitem [{\citenamefont {Morgan}\ and\ \citenamefont
  {Watson}(2009)}]{morgan2009polaronic}%
  \BibitemOpen
  \bibfield  {author} {\bibinfo {author} {\bibfnamefont {B.~J.}\ \bibnamefont
  {Morgan}}\ and\ \bibinfo {author} {\bibfnamefont {G.~W.}\ \bibnamefont
  {Watson}},\ }\href@noop {} {\bibfield  {journal} {\bibinfo  {journal} {Phys.
  Rev. B}\ }\textbf {\bibinfo {volume} {80}},\ \bibinfo {pages} {233102}
  (\bibinfo {year} {2009})}\BibitemShut {NoStop}%
\bibitem [{\citenamefont {Sokol}, \citenamefont {Walsh},\ and\ \citenamefont
  {Catlow}(2010)}]{sokol2010oxygen}%
  \BibitemOpen
  \bibfield  {author} {\bibinfo {author} {\bibfnamefont {A.~A.}\ \bibnamefont
  {Sokol}}, \bibinfo {author} {\bibfnamefont {A.}~\bibnamefont {Walsh}}, \ and\
  \bibinfo {author} {\bibfnamefont {C.~R.~A.}\ \bibnamefont {Catlow}},\
  }\href@noop {} {\bibfield  {journal} {\bibinfo  {journal} {Chem. Phys.
  Lett.}\ }\textbf {\bibinfo {volume} {492}},\ \bibinfo {pages} {44} (\bibinfo
  {year} {2010})}\BibitemShut {NoStop}%
\bibitem [{\citenamefont {Zhang}\ and\ \citenamefont
  {Northrup}(1991)}]{zhang1991chemical}%
  \BibitemOpen
  \bibfield  {author} {\bibinfo {author} {\bibfnamefont {S.}~\bibnamefont
  {Zhang}}\ and\ \bibinfo {author} {\bibfnamefont {J.~E.}\ \bibnamefont
  {Northrup}},\ }\href@noop {} {\bibfield  {journal} {\bibinfo  {journal}
  {Phys. Rev. Lett.}\ }\textbf {\bibinfo {volume} {67}},\ \bibinfo {pages}
  {2339} (\bibinfo {year} {1991})}\BibitemShut {NoStop}%
\bibitem [{\citenamefont {Buckeridge}\ \emph {et~al.}(2014)\citenamefont
  {Buckeridge}, \citenamefont {Scanlon}, \citenamefont {Walsh},\ and\
  \citenamefont {Catlow}}]{buckeridge2014automated}%
  \BibitemOpen
  \bibfield  {author} {\bibinfo {author} {\bibfnamefont {J.}~\bibnamefont
  {Buckeridge}}, \bibinfo {author} {\bibfnamefont {D.~O.}\ \bibnamefont
  {Scanlon}}, \bibinfo {author} {\bibfnamefont {A.}~\bibnamefont {Walsh}}, \
  and\ \bibinfo {author} {\bibfnamefont {C.~R.~A.}\ \bibnamefont {Catlow}},\
  }\href@noop {} {\bibfield  {journal} {\bibinfo  {journal} {Comp. Phys.
  Commun.}\ }\textbf {\bibinfo {volume} {185}},\ \bibinfo {pages} {330}
  (\bibinfo {year} {2014})}\BibitemShut {NoStop}%
\bibitem [{\citenamefont {Aradi}\ \emph {et~al.}(2001)\citenamefont {Aradi},
  \citenamefont {Gali}, \citenamefont {De{\'a}k}, \citenamefont {Lowther},
  \citenamefont {Son}, \citenamefont {Janz{\'e}n},\ and\ \citenamefont
  {Choyke}}]{aradi2001ab}%
  \BibitemOpen
  \bibfield  {author} {\bibinfo {author} {\bibfnamefont {B.}~\bibnamefont
  {Aradi}}, \bibinfo {author} {\bibfnamefont {A.}~\bibnamefont {Gali}},
  \bibinfo {author} {\bibfnamefont {P.}~\bibnamefont {De{\'a}k}}, \bibinfo
  {author} {\bibfnamefont {J.}~\bibnamefont {Lowther}}, \bibinfo {author}
  {\bibfnamefont {N.}~\bibnamefont {Son}}, \bibinfo {author} {\bibfnamefont
  {E.}~\bibnamefont {Janz{\'e}n}}, \ and\ \bibinfo {author} {\bibfnamefont
  {W.}~\bibnamefont {Choyke}},\ }\href@noop {} {\bibfield  {journal} {\bibinfo
  {journal} {Phy. Rev. B}\ }\textbf {\bibinfo {volume} {63}},\ \bibinfo {pages}
  {245202} (\bibinfo {year} {2001})}\BibitemShut {NoStop}%
\bibitem [{\citenamefont {Buckeridge}(2019)}]{buckeridge2019equilibrium}%
  \BibitemOpen
  \bibfield  {author} {\bibinfo {author} {\bibfnamefont {J.}~\bibnamefont
  {Buckeridge}},\ }\href@noop {} {\bibfield  {journal} {\bibinfo  {journal}
  {Comp. Phys. Commun.}\ }\textbf {\bibinfo {volume} {244}},\ \bibinfo {pages}
  {329} (\bibinfo {year} {2019})}\BibitemShut {NoStop}%
\bibitem [{\citenamefont {Naik}\ and\ \citenamefont {Jain}(2018)}]{Naik2018}%
  \BibitemOpen
  \bibfield  {author} {\bibinfo {author} {\bibfnamefont {M.~H.}\ \bibnamefont
  {Naik}}\ and\ \bibinfo {author} {\bibfnamefont {M.}~\bibnamefont {Jain}},\
  }\href {\doibase https://doi.org/10.1016/j.cpc.2018.01.011} {\bibfield
  {journal} {\bibinfo  {journal} {Comput. Phys. Commun.}\ }\textbf {\bibinfo
  {volume} {226}},\ \bibinfo {pages} {114 } (\bibinfo {year}
  {2018})}\BibitemShut {NoStop}%
\bibitem [{\citenamefont {Péan}\ \emph {et~al.}(2017)\citenamefont {Péan},
  \citenamefont {Vidal}, \citenamefont {Jobic},\ and\ \citenamefont
  {Latouche}}]{Pean2017}%
  \BibitemOpen
  \bibfield  {author} {\bibinfo {author} {\bibfnamefont {E.}~\bibnamefont
  {Péan}}, \bibinfo {author} {\bibfnamefont {J.}~\bibnamefont {Vidal}},
  \bibinfo {author} {\bibfnamefont {S.}~\bibnamefont {Jobic}}, \ and\ \bibinfo
  {author} {\bibfnamefont {C.}~\bibnamefont {Latouche}},\ }\href {\doibase
  https://doi.org/10.1016/j.cplett.2017.01.001} {\bibfield  {journal} {\bibinfo
   {journal} {Chem. Phys. Lett.}\ }\textbf {\bibinfo {volume} {671}},\ \bibinfo
  {pages} {124 } (\bibinfo {year} {2017})}\BibitemShut {NoStop}%
\bibitem [{\citenamefont {Freysoldt}, \citenamefont {Neugebauer},\ and\
  \citenamefont {Van~de Walle}(2009{\natexlab{a}})}]{FNV2009}%
  \BibitemOpen
  \bibfield  {author} {\bibinfo {author} {\bibfnamefont {C.}~\bibnamefont
  {Freysoldt}}, \bibinfo {author} {\bibfnamefont {J.}~\bibnamefont
  {Neugebauer}}, \ and\ \bibinfo {author} {\bibfnamefont {C.~G.}\ \bibnamefont
  {Van~de Walle}},\ }\href {\doibase 10.1103/PhysRevLett.102.016402} {\bibfield
   {journal} {\bibinfo  {journal} {Phys. Rev. Lett.}\ }\textbf {\bibinfo
  {volume} {102}},\ \bibinfo {pages} {016402} (\bibinfo {year}
  {2009}{\natexlab{a}})}\BibitemShut {NoStop}%
\bibitem [{\citenamefont {Marquardt}\ \emph {et~al.}(2012)\citenamefont
  {Marquardt}, \citenamefont {Schulz}, \citenamefont {Freysoldt}, \citenamefont
  {Boeck}, \citenamefont {Hickel}, \citenamefont {O'Reilly},\ and\
  \citenamefont {Neugebauer}}]{Marquardt2012}%
  \BibitemOpen
  \bibfield  {author} {\bibinfo {author} {\bibfnamefont {O.}~\bibnamefont
  {Marquardt}}, \bibinfo {author} {\bibfnamefont {S.}~\bibnamefont {Schulz}},
  \bibinfo {author} {\bibfnamefont {C.}~\bibnamefont {Freysoldt}}, \bibinfo
  {author} {\bibfnamefont {S.}~\bibnamefont {Boeck}}, \bibinfo {author}
  {\bibfnamefont {T.}~\bibnamefont {Hickel}}, \bibinfo {author} {\bibfnamefont
  {E.~P.}\ \bibnamefont {O'Reilly}}, \ and\ \bibinfo {author} {\bibfnamefont
  {J.}~\bibnamefont {Neugebauer}},\ }\href {\doibase 10.1007/s11082-011-9506-3}
  {\bibfield  {journal} {\bibinfo  {journal} {Opt. and Quant. Electron.}\
  }\textbf {\bibinfo {volume} {44}},\ \bibinfo {pages} {183} (\bibinfo {year}
  {2012})}\BibitemShut {NoStop}%
\bibitem [{\citenamefont {Freysoldt}\ and\ \citenamefont
  {Neugebauer}(2018{\natexlab{a}})}]{Freysoldt2018interface}%
  \BibitemOpen
  \bibfield  {author} {\bibinfo {author} {\bibfnamefont {C.}~\bibnamefont
  {Freysoldt}}\ and\ \bibinfo {author} {\bibfnamefont {J.}~\bibnamefont
  {Neugebauer}},\ }\href {\doibase 10.1103/PhysRevB.97.205425} {\bibfield
  {journal} {\bibinfo  {journal} {Phys. Rev. B}\ }\textbf {\bibinfo {volume}
  {97}},\ \bibinfo {pages} {205425} (\bibinfo {year}
  {2018}{\natexlab{a}})}\BibitemShut {NoStop}%
\bibitem [{\citenamefont {Kim}\ \emph {et~al.}(2020{\natexlab{a}})\citenamefont
  {Kim}, \citenamefont {Hood}, \citenamefont {van Gerwen}, \citenamefont
  {Whalley},\ and\ \citenamefont {Walsh}}]{Kim:2020bx}%
  \BibitemOpen
  \bibfield  {author} {\bibinfo {author} {\bibfnamefont {S.}~\bibnamefont
  {Kim}}, \bibinfo {author} {\bibfnamefont {S.}~\bibnamefont {Hood}}, \bibinfo
  {author} {\bibfnamefont {P.}~\bibnamefont {van Gerwen}}, \bibinfo {author}
  {\bibfnamefont {L.}~\bibnamefont {Whalley}}, \ and\ \bibinfo {author}
  {\bibfnamefont {A.}~\bibnamefont {Walsh}},\ }\href@noop {} {\bibfield
  {journal} {\bibinfo  {journal} {J. Open Source Softw.}\ }\textbf {\bibinfo
  {volume} {5}},\ \bibinfo {pages} {2102} (\bibinfo {year}
  {2020}{\natexlab{a}})}\BibitemShut {NoStop}%
\bibitem [{\citenamefont {Durrant}\ \emph {et~al.}(2018)\citenamefont
  {Durrant}, \citenamefont {Murphy}, \citenamefont {Watkins},\ and\
  \citenamefont {Shluger}}]{durrant2018}%
  \BibitemOpen
  \bibfield  {author} {\bibinfo {author} {\bibfnamefont {T.~R.}\ \bibnamefont
  {Durrant}}, \bibinfo {author} {\bibfnamefont {S.~T.}\ \bibnamefont {Murphy}},
  \bibinfo {author} {\bibfnamefont {M.~B.}\ \bibnamefont {Watkins}}, \ and\
  \bibinfo {author} {\bibfnamefont {A.~L.}\ \bibnamefont {Shluger}},\ }\href
  {\doibase 10.1063/1.5029818} {\bibfield  {journal} {\bibinfo  {journal} {J.
  Chem. Phys.}\ }\textbf {\bibinfo {volume} {149}},\ \bibinfo {pages} {024103}
  (\bibinfo {year} {2018})}\BibitemShut {NoStop}%
\bibitem [{\citenamefont {Vinichenko}\ \emph {et~al.}(2017)\citenamefont
  {Vinichenko}, \citenamefont {Sensoy}, \citenamefont {Friend},\ and\
  \citenamefont {Kaxiras}}]{Vinichenko2017}%
  \BibitemOpen
  \bibfield  {author} {\bibinfo {author} {\bibfnamefont {D.}~\bibnamefont
  {Vinichenko}}, \bibinfo {author} {\bibfnamefont {M.~G.}\ \bibnamefont
  {Sensoy}}, \bibinfo {author} {\bibfnamefont {C.~M.}\ \bibnamefont {Friend}},
  \ and\ \bibinfo {author} {\bibfnamefont {E.}~\bibnamefont {Kaxiras}},\ }\href
  {\doibase 10.1103/PhysRevB.95.235310} {\bibfield  {journal} {\bibinfo
  {journal} {Phys. Rev. B}\ }\textbf {\bibinfo {volume} {95}},\ \bibinfo
  {pages} {235310} (\bibinfo {year} {2017})}\BibitemShut {NoStop}%
\bibitem [{\citenamefont {Leslie}\ and\ \citenamefont
  {Gillan}(1985)}]{Leslie1985}%
  \BibitemOpen
  \bibfield  {author} {\bibinfo {author} {\bibfnamefont {M.}~\bibnamefont
  {Leslie}}\ and\ \bibinfo {author} {\bibfnamefont {M.~J.}\ \bibnamefont
  {Gillan}},\ }\href {\doibase 10.1088/0022-3719/18/5/005} {\bibfield
  {journal} {\bibinfo  {journal} {J. Phys. C: Solid State Phys.}\ }\textbf
  {\bibinfo {volume} {18}},\ \bibinfo {pages} {973} (\bibinfo {year}
  {1985})}\BibitemShut {NoStop}%
\bibitem [{\citenamefont {Makov}\ and\ \citenamefont
  {Payne}(1995)}]{Makov1995}%
  \BibitemOpen
  \bibfield  {author} {\bibinfo {author} {\bibfnamefont {G.}~\bibnamefont
  {Makov}}\ and\ \bibinfo {author} {\bibfnamefont {M.~C.}\ \bibnamefont
  {Payne}},\ }\href {\doibase 10.1103/PhysRevB.51.4014} {\bibfield  {journal}
  {\bibinfo  {journal} {Phys. Rev. B}\ }\textbf {\bibinfo {volume} {51}},\
  \bibinfo {pages} {4014} (\bibinfo {year} {1995})}\BibitemShut {NoStop}%
\bibitem [{\citenamefont {Lany}\ and\ \citenamefont {Zunger}(2009)}]{Lany2009}%
  \BibitemOpen
  \bibfield  {author} {\bibinfo {author} {\bibfnamefont {S.}~\bibnamefont
  {Lany}}\ and\ \bibinfo {author} {\bibfnamefont {A.}~\bibnamefont {Zunger}},\
  }\href {\doibase 10.1088/0965-0393/17/8/084002} {\bibfield  {journal}
  {\bibinfo  {journal} {Modell. Simul. Mater. Sci. Eng.}\ }\textbf {\bibinfo
  {volume} {17}},\ \bibinfo {pages} {084002} (\bibinfo {year}
  {2009})}\BibitemShut {NoStop}%
\bibitem [{\citenamefont {Freysoldt}, \citenamefont {Neugebauer},\ and\
  \citenamefont {Van~de Walle}(2009{\natexlab{b}})}]{Freysoldt2009}%
  \BibitemOpen
  \bibfield  {author} {\bibinfo {author} {\bibfnamefont {C.}~\bibnamefont
  {Freysoldt}}, \bibinfo {author} {\bibfnamefont {J.}~\bibnamefont
  {Neugebauer}}, \ and\ \bibinfo {author} {\bibfnamefont {C.~G.}\ \bibnamefont
  {Van~de Walle}},\ }\href {\doibase 10.1103/PhysRevLett.102.016402} {\bibfield
   {journal} {\bibinfo  {journal} {Phys. Rev. Lett.}\ }\textbf {\bibinfo
  {volume} {102}},\ \bibinfo {pages} {016402} (\bibinfo {year}
  {2009}{\natexlab{b}})}\BibitemShut {NoStop}%
\bibitem [{\citenamefont {Kumagai}\ and\ \citenamefont
  {Oba}(2014)}]{Kumagai2014}%
  \BibitemOpen
  \bibfield  {author} {\bibinfo {author} {\bibfnamefont {Y.}~\bibnamefont
  {Kumagai}}\ and\ \bibinfo {author} {\bibfnamefont {F.}~\bibnamefont {Oba}},\
  }\href {\doibase 10.1103/PhysRevB.89.195205} {\bibfield  {journal} {\bibinfo
  {journal} {Phys. Rev. B}\ }\textbf {\bibinfo {volume} {89}},\ \bibinfo
  {pages} {195205} (\bibinfo {year} {2014})}\BibitemShut {NoStop}%
\bibitem [{\citenamefont {Freysoldt}\ and\ \citenamefont
  {Neugebauer}(2018{\natexlab{b}})}]{Freysoldt2018}%
  \BibitemOpen
  \bibfield  {author} {\bibinfo {author} {\bibfnamefont {C.}~\bibnamefont
  {Freysoldt}}\ and\ \bibinfo {author} {\bibfnamefont {J.}~\bibnamefont
  {Neugebauer}},\ }\href {\doibase 10.1103/PhysRevB.97.205425} {\bibfield
  {journal} {\bibinfo  {journal} {Phys. Rev. B}\ }\textbf {\bibinfo {volume}
  {97}},\ \bibinfo {pages} {205425} (\bibinfo {year}
  {2018}{\natexlab{b}})}\BibitemShut {NoStop}%
\bibitem [{\citenamefont {Komsa}\ and\ \citenamefont
  {Pasquarello}(2013)}]{Komsa2013}%
  \BibitemOpen
  \bibfield  {author} {\bibinfo {author} {\bibfnamefont {H.-P.}\ \bibnamefont
  {Komsa}}\ and\ \bibinfo {author} {\bibfnamefont {A.}~\bibnamefont
  {Pasquarello}},\ }\href {\doibase 10.1103/PhysRevLett.110.095505} {\bibfield
  {journal} {\bibinfo  {journal} {Phys. Rev. Lett.}\ }\textbf {\bibinfo
  {volume} {110}},\ \bibinfo {pages} {095505} (\bibinfo {year}
  {2013})}\BibitemShut {NoStop}%
\bibitem [{\citenamefont {Kim}, \citenamefont {Chang},\ and\ \citenamefont
  {Park}(2014)}]{Kim2014}%
  \BibitemOpen
  \bibfield  {author} {\bibinfo {author} {\bibfnamefont {S.}~\bibnamefont
  {Kim}}, \bibinfo {author} {\bibfnamefont {K.~J.}\ \bibnamefont {Chang}}, \
  and\ \bibinfo {author} {\bibfnamefont {J.-S.}\ \bibnamefont {Park}},\ }\href
  {\doibase 10.1103/PhysRevB.90.085435} {\bibfield  {journal} {\bibinfo
  {journal} {Phys. Rev. B}\ }\textbf {\bibinfo {volume} {90}},\ \bibinfo
  {pages} {085435} (\bibinfo {year} {2014})}\BibitemShut {NoStop}%
\bibitem [{\citenamefont {Lany}\ and\ \citenamefont
  {Zunger}(2008)}]{Lany:2008gk}%
  \BibitemOpen
  \bibfield  {author} {\bibinfo {author} {\bibfnamefont {S.}~\bibnamefont
  {Lany}}\ and\ \bibinfo {author} {\bibfnamefont {A.}~\bibnamefont {Zunger}},\
  }\href@noop {} {\bibfield  {journal} {\bibinfo  {journal} {Phys. Rev. B}\
  }\textbf {\bibinfo {volume} {78}},\ \bibinfo {pages} {2637} (\bibinfo {year}
  {2008})}\BibitemShut {NoStop}%
\bibitem [{\citenamefont {Shi}\ and\ \citenamefont {Wang}(2012)}]{Shi:2012iv}%
  \BibitemOpen
  \bibfield  {author} {\bibinfo {author} {\bibfnamefont {L.}~\bibnamefont
  {Shi}}\ and\ \bibinfo {author} {\bibfnamefont {L.-W.}\ \bibnamefont {Wang}},\
  }\href@noop {} {\bibfield  {journal} {\bibinfo  {journal} {Phys. Rev. Lett.}\
  }\textbf {\bibinfo {volume} {109}},\ \bibinfo {pages} {245501} (\bibinfo
  {year} {2012})}\BibitemShut {NoStop}%
\bibitem [{\citenamefont {Alkauskas}, \citenamefont {Yan},\ and\ \citenamefont
  {Van~de Walle}(2014)}]{Alkauskas:2014kk}%
  \BibitemOpen
  \bibfield  {author} {\bibinfo {author} {\bibfnamefont {A.}~\bibnamefont
  {Alkauskas}}, \bibinfo {author} {\bibfnamefont {Q.}~\bibnamefont {Yan}}, \
  and\ \bibinfo {author} {\bibfnamefont {C.~G.}\ \bibnamefont {Van~de Walle}},\
  }\href@noop {} {\bibfield  {journal} {\bibinfo  {journal} {Phys. Rev. B}\
  }\textbf {\bibinfo {volume} {90}},\ \bibinfo {pages} {075202} (\bibinfo
  {year} {2014})}\BibitemShut {NoStop}%
\bibitem [{\citenamefont {Kim}, \citenamefont {Hood},\ and\ \citenamefont
  {Walsh}(2019)}]{Kim:2019hp}%
  \BibitemOpen
  \bibfield  {author} {\bibinfo {author} {\bibfnamefont {S.}~\bibnamefont
  {Kim}}, \bibinfo {author} {\bibfnamefont {S.~N.}\ \bibnamefont {Hood}}, \
  and\ \bibinfo {author} {\bibfnamefont {A.}~\bibnamefont {Walsh}},\
  }\href@noop {} {\bibfield  {journal} {\bibinfo  {journal} {Phys. Rev. B}\
  }\textbf {\bibinfo {volume} {100}},\ \bibinfo {pages} {041202} (\bibinfo
  {year} {2019})}\BibitemShut {NoStop}%
\bibitem [{\citenamefont {Kim}\ \emph {et~al.}(2020{\natexlab{b}})\citenamefont
  {Kim}, \citenamefont {Marquez}, \citenamefont {Unold},\ and\ \citenamefont
  {Walsh}}]{Kim:2020sc}%
  \BibitemOpen
  \bibfield  {author} {\bibinfo {author} {\bibfnamefont {S.}~\bibnamefont
  {Kim}}, \bibinfo {author} {\bibfnamefont {J.~A.}\ \bibnamefont {Marquez}},
  \bibinfo {author} {\bibfnamefont {T.}~\bibnamefont {Unold}}, \ and\ \bibinfo
  {author} {\bibfnamefont {A.}~\bibnamefont {Walsh}},\ }\href@noop {}
  {\bibfield  {journal} {\bibinfo  {journal} {Energ. Environ. Sci.}\ ,\
  \bibinfo {pages} {Advance article}} (\bibinfo {year}
  {2020}{\natexlab{b}})}\BibitemShut {NoStop}%
\bibitem [{\citenamefont {Vineyard}(1957)}]{Vineyard:1957iy}%
  \BibitemOpen
  \bibfield  {author} {\bibinfo {author} {\bibfnamefont {G.~H.}\ \bibnamefont
  {Vineyard}},\ }\href@noop {} {\bibfield  {journal} {\bibinfo  {journal} {J.
  Phys. Chem. Solids}\ }\textbf {\bibinfo {volume} {3}},\ \bibinfo {pages}
  {121} (\bibinfo {year} {1957})}\BibitemShut {NoStop}%
\bibitem [{\citenamefont {Henkelman}, \citenamefont {Uberuaga},\ and\
  \citenamefont {J{\'o}nsson}(2000)}]{Henkelman:2000ez}%
  \BibitemOpen
  \bibfield  {author} {\bibinfo {author} {\bibfnamefont {G.}~\bibnamefont
  {Henkelman}}, \bibinfo {author} {\bibfnamefont {B.~P.}\ \bibnamefont
  {Uberuaga}}, \ and\ \bibinfo {author} {\bibfnamefont {H.}~\bibnamefont
  {J{\'o}nsson}},\ }\href@noop {} {\bibfield  {journal} {\bibinfo  {journal}
  {J. Chem. Phys.}\ }\textbf {\bibinfo {volume} {113}},\ \bibinfo {pages}
  {9901} (\bibinfo {year} {2000})}\BibitemShut {NoStop}%
\bibitem [{\citenamefont {Yang}\ \emph {et~al.}(2015)\citenamefont {Yang},
  \citenamefont {Park}, \citenamefont {Kang},\ and\ \citenamefont
  {Wei}}]{yang2015first}%
  \BibitemOpen
  \bibfield  {author} {\bibinfo {author} {\bibfnamefont {J.-H.}\ \bibnamefont
  {Yang}}, \bibinfo {author} {\bibfnamefont {J.-S.}\ \bibnamefont {Park}},
  \bibinfo {author} {\bibfnamefont {J.}~\bibnamefont {Kang}}, \ and\ \bibinfo
  {author} {\bibfnamefont {S.-H.}\ \bibnamefont {Wei}},\ }\href@noop {}
  {\bibfield  {journal} {\bibinfo  {journal} {Phys. Rev. B}\ }\textbf {\bibinfo
  {volume} {91}},\ \bibinfo {pages} {075202} (\bibinfo {year}
  {2015})}\BibitemShut {NoStop}%
\bibitem [{\citenamefont {Kim}\ \emph {et~al.}(2018)\citenamefont {Kim},
  \citenamefont {Senocrate}, \citenamefont {Yang}, \citenamefont {Gregori},
  \citenamefont {Gr{\"a}tzel},\ and\ \citenamefont {Maier}}]{kim2018large}%
  \BibitemOpen
  \bibfield  {author} {\bibinfo {author} {\bibfnamefont {G.~Y.}\ \bibnamefont
  {Kim}}, \bibinfo {author} {\bibfnamefont {A.}~\bibnamefont {Senocrate}},
  \bibinfo {author} {\bibfnamefont {T.-Y.}\ \bibnamefont {Yang}}, \bibinfo
  {author} {\bibfnamefont {G.}~\bibnamefont {Gregori}}, \bibinfo {author}
  {\bibfnamefont {M.}~\bibnamefont {Gr{\"a}tzel}}, \ and\ \bibinfo {author}
  {\bibfnamefont {J.}~\bibnamefont {Maier}},\ }\href@noop {} {\bibfield
  {journal} {\bibinfo  {journal} {Nat. Mater.}\ }\textbf {\bibinfo {volume}
  {17}},\ \bibinfo {pages} {445} (\bibinfo {year} {2018})}\BibitemShut
  {NoStop}%
\bibitem [{\citenamefont {Walsh}\ and\ \citenamefont
  {Stranks}(2018)}]{walsh2018taking}%
  \BibitemOpen
  \bibfield  {author} {\bibinfo {author} {\bibfnamefont {A.}~\bibnamefont
  {Walsh}}\ and\ \bibinfo {author} {\bibfnamefont {S.~D.}\ \bibnamefont
  {Stranks}},\ }\href@noop {} {\bibfield  {journal} {\bibinfo  {journal} {ACS
  Energy Lett.}\ }\textbf {\bibinfo {volume} {3}},\ \bibinfo {pages} {1983}
  (\bibinfo {year} {2018})}\BibitemShut {NoStop}%
\bibitem [{\citenamefont {Pizzi}\ \emph {et~al.}(2016)\citenamefont {Pizzi},
  \citenamefont {Cepellotti}, \citenamefont {Sabatini}, \citenamefont
  {Marzari},\ and\ \citenamefont {Kozinsky}}]{pizzi2016aiida}%
  \BibitemOpen
  \bibfield  {author} {\bibinfo {author} {\bibfnamefont {G.}~\bibnamefont
  {Pizzi}}, \bibinfo {author} {\bibfnamefont {A.}~\bibnamefont {Cepellotti}},
  \bibinfo {author} {\bibfnamefont {R.}~\bibnamefont {Sabatini}}, \bibinfo
  {author} {\bibfnamefont {N.}~\bibnamefont {Marzari}}, \ and\ \bibinfo
  {author} {\bibfnamefont {B.}~\bibnamefont {Kozinsky}},\ }\href@noop {}
  {\bibfield  {journal} {\bibinfo  {journal} {Comp. Mater. Sci.}\ }\textbf
  {\bibinfo {volume} {111}},\ \bibinfo {pages} {218} (\bibinfo {year}
  {2016})}\BibitemShut {NoStop}%
\bibitem [{\citenamefont {Mannodi-Kanakkithodi}\ \emph
  {et~al.}(2020)\citenamefont {Mannodi-Kanakkithodi}, \citenamefont {Toriyama},
  \citenamefont {Sen}, \citenamefont {Davis}, \citenamefont {Klie},\ and\
  \citenamefont {Chan}}]{mannodi2020machine}%
  \BibitemOpen
  \bibfield  {author} {\bibinfo {author} {\bibfnamefont {A.}~\bibnamefont
  {Mannodi-Kanakkithodi}}, \bibinfo {author} {\bibfnamefont {M.~Y.}\
  \bibnamefont {Toriyama}}, \bibinfo {author} {\bibfnamefont {F.~G.}\
  \bibnamefont {Sen}}, \bibinfo {author} {\bibfnamefont {M.~J.}\ \bibnamefont
  {Davis}}, \bibinfo {author} {\bibfnamefont {R.~F.}\ \bibnamefont {Klie}}, \
  and\ \bibinfo {author} {\bibfnamefont {M.~K.}\ \bibnamefont {Chan}},\
  }\href@noop {} {\bibfield  {journal} {\bibinfo  {journal} {npj Comp. Mater.}\
  }\textbf {\bibinfo {volume} {6}},\ \bibinfo {pages} {1} (\bibinfo {year}
  {2020})}\BibitemShut {NoStop}%
\end{thebibliography}%

\end{document}